\documentclass[conference]{IEEEtran}
\IEEEoverridecommandlockouts
\usepackage{amsmath,amssymb,amsfonts}
\usepackage{algorithmic}
\usepackage{textcomp}
\usepackage{xcolor}
\usepackage{tabularx}
\usepackage{epsfig}
\usepackage{multicol}
\usepackage{booktabs}
\usepackage{pslatex}
\usepackage{url}

\usepackage[
backend=biber,
style=numeric,
sorting=ynt
]{biblatex}
\begin{document}

\title{The AffectToolbox: Affect Analysis for Everyone
}

\author{\IEEEauthorblockN{Silvan Mertes, Dominik Schiller, Michael Dietz, Elisabeth André and Florian Lingenfelser}
\IEEEauthorblockA{\textit{Human-Centered Artificial Intelligence, Augsburg University} \\
Augsburg, Germany \\
\{forename.surname\}@uni-a.de}

}

\maketitle

\begin{abstract}
In the field of affective computing, where research continually advances at a rapid pace, the demand for user-friendly tools has become increasingly apparent. In this paper, we present the \emph{AffectToolbox}, a novel software system that aims to support researchers in developing affect-sensitive studies and prototypes. 
The proposed system addresses the challenges posed by existing frameworks, which often require profound programming knowledge and cater primarily to power-users or skilled developers. 
Aiming to facilitate ease of use, the \emph{AffectToolbox} requires no programming knowledge and offers its functionality to reliably analyze the affective state of users through an accessible graphical user interface. The architecture encompasses a variety of models for emotion recognition on multiple affective channels and modalities, as well as an elaborate fusion system to merge multi-modal assessments into a unified result.
The entire system is open-sourced and will be publicly available to ensure easy integration into more complex applications through a well-structured, Python-based code base - therefore marking a substantial contribution toward advancing affective computing research and fostering a more collaborative and inclusive environment within this interdisciplinary field.
\end{abstract}


\section{Introduction}
\label{sec:intro}
\emph{Affective computing} describes a field of research, which incorporates the recognition, interpretation and simulation of human emotions in software systems. In particular, the \emph{AffectToolbox} aims at the automated recognition of affective states - a sub-discipline of \emph{machine learning} and \emph{artificial intelligence} research. As part of this ever-evolving realm, affect recognition is characterized by rapid pace of new developments and a continuous pushing of boundaries in research.
% In the ever-evolving realm of affective computing, the research landscape is characterized by its rapid pace, continually pushing the boundaries of research and development. 
%The exploration of affective computing often involves addressing recurring tasks such as estimating a user's valence, arousal, and dominance, while employing various modalities like speech prosody, sentiment, facial expressions, and pose.
The exploration of affective computing often involves addressing recurring tasks such as interpreting facial expressions, analysing speech prosody and sentiment or observing non-verbal behaviour in body movements and poses. 
However, despite the growing interest and demand in this field, existing tools and frameworks (e.g. Microsoft's \emph{Platform for Situated Intelligence}\cite{bohus2021platform} or Google's \emph{MediaPipe}\cite{lugaresi2019mediapipe}) present challenges for swift development, as they require in-depth programming knowledge and are primarily aimed at power users or experienced developers.

Even within more accessible systems, a discernible gap exists, as most of them are focusing predominantly on providing the input modalities and signals, while lacking essential built-in functionality for comprehensive affect analysis and recognition. 
This limitation underscores the need for more versatile, accessible, and comprehensive architectures that cater to a broader audience, including researchers and practitioners without extensive programming backgrounds but high interest and demand for most recent solutions. 
Addressing these challenges is pivotal to advancing affective computing research and fostering a more inclusive and collaborative environment in this dynamic and interdisciplinary field.\newline

To close this gap, in this paper, we present the \emph{AffectToolbox}, a comprehensive affect recognition software system,\\ \textbf{that is}
\begin{itemize}
    \item \textbf{Easy to use.} No programming knowledge required - all necessary functionality can easily be controlled by a graphical user interface.
    \item \textbf{Comprehensive.} A variety of models for interpreting accessible affective channels is included.
    % \item \textbf{easy to integrate in applications.} The AffectToolbox can easily be integrated into more complex applications, as it is open-sourced as Python framework. Further, it has broad networking functionality included, allowing it to seamlessly being connected to proprietary software implementations.
    \item \textbf{Easy to integrate.} The \emph{AffectToolbox} can easily be integrated into more complex applications, using its built-in networking functionality, allowing for a seamless connection to proprietary software implementations.
\item \textbf{Open source.} The whole software system and source code will be made publicly available.\footnote{\url{https://github.com/hcmlab/AffectToolbox}}
\end{itemize}

\begin{figure*}[t]
\centering
\includegraphics[width=0.9\textwidth]{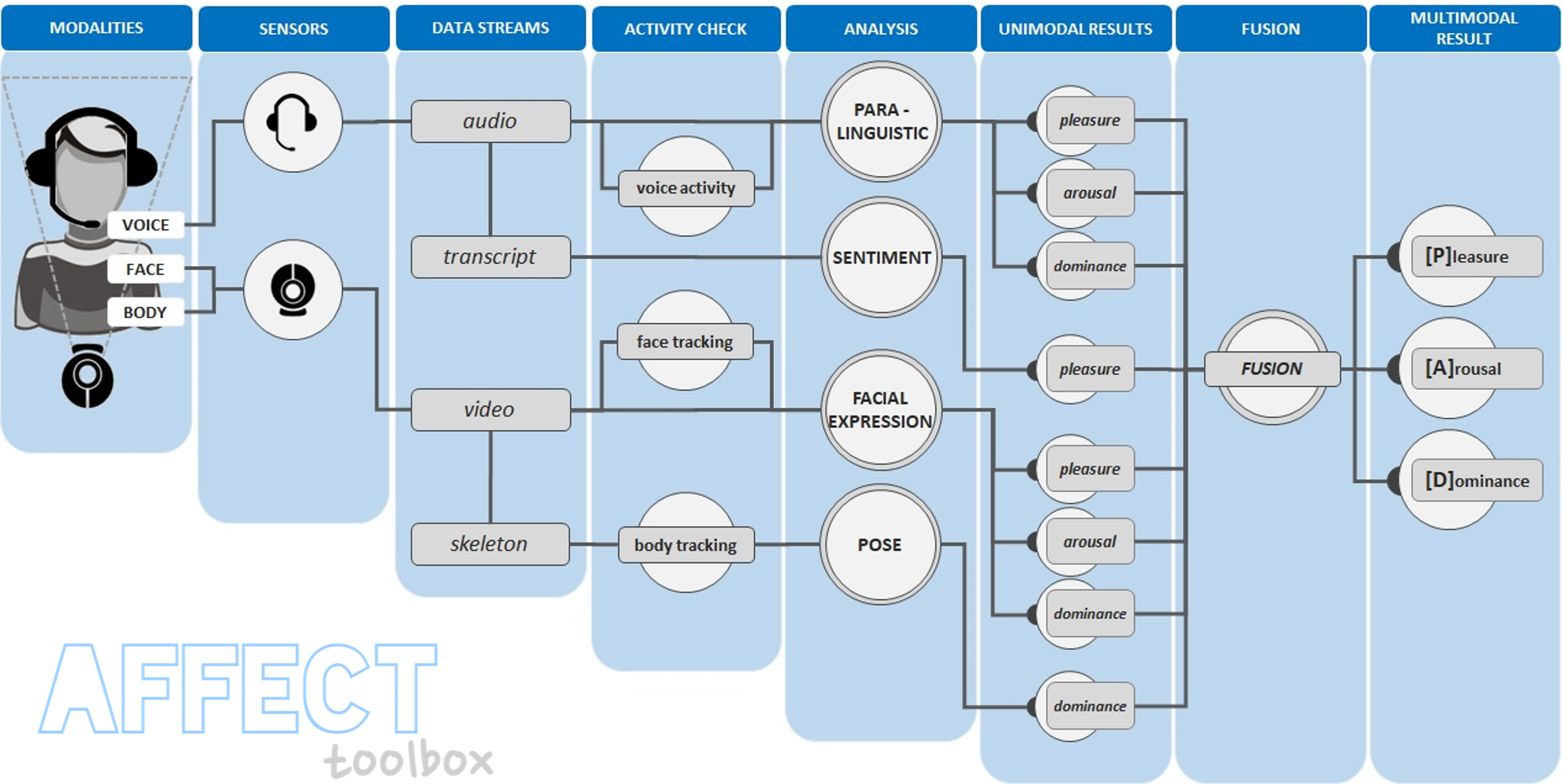}
\caption{The modular architecture of the \emph{AffectToolbox} (Section \ref{sec:system}). In its current state, audiovisual sensory devices (e.g. webcams) provide easy means to generate all considered data streams (i.e. audio, transcript, video and skeleton data). So called activity checks trigger the machine learning based analysis of respective modalities (Section \ref{sec:modalities}). The uni-modal results of applied affect recognition models are represented by a subset of pleasure, arousal and/or dominance scores. These unimodal emotional cues are the input for an event-driven fusion algorithm (Section \ref{sec:fusion}), which deduces a coherent affective state, represented in the continuous PAD emotional space (Section \ref{sec:emo}).}
\label{fig:system}
\end{figure*}

\section{Related Work}
\label{sec:related}
To give a foundation for the conceptualization and development of the \emph{AffectToolbox}, in the following we give a brief overview on existing software that addresses related goals.

\subsection{Multi-modal Sensor Integration and Synchronization}
\emph{Apache Flink} \cite{carbone2015apache} is an open-source stream processing and batch processing framework for big data processing and analytics. It is designed to efficiently process large volumes of data in real-time and batch processing modes. \emph{Flink} provides a programming model and runtime for the processing of data streams in a distributed and fault-tolerant manner.

\emph{SiAM-dp} \cite{nesselrath2015siam} serves as a platform dedicated to the development of multi-modal dialogue systems, with a key emphasis on effortlessly integrating dispersed input and output devices within the realm of cyber-physical environments. 

The \emph{Social Signal Interpretation Framework} (\emph{SSI}) \cite{wagner2013social} is an efficient framework that offers robust support for diverse sensor devices, filter and feature algorithms, and provides an extensive C++ API to add further functionality and train custom machine learning models. Meanwhile, for end users, \emph{SSI} provides accessibility through an XML interface. To also be accessible on mobile devices, Damian et al. ported the core functionality of \emph{SSI} to a Java-based framework which they called \emph{SSJ} \cite{damian2018ssj}.

Barz et al. introduced the \emph{Multisensor Pipeline} (\emph{MSP}) \cite{barz2021multisensor}, a Python framework that serves as lightweight tool for prototyping multi-modal sensor pipelines. A similar goal was followed by Saffaryazdi et al. \cite{saffaryazdi2022octopus}, who introduced a Python framework called \emph{Octopus Sensing} - however, in contrast to \emph{MSP}, which focuses more on real-time applications, \emph{Octopus Sensing} specifically addresses multi-modal data collection.

To the best of the authors' knowledge, all of the frameworks mentioned above, although providing comprehensive functionality to track and synchronize a multitude of different sensor hardware, do not  provide pre-trained machine learning models, which allow for an immediate in-depth affective analysis of the modalities. They are limited to providing an efficient and extensible infrastructure for further computations (which may include the definition and training of custom classification models).

\subsection{Multi-modal Signal Analysis}
Due to the ongoing growth of the field, some large companies have also recognised the economic potential of according systems recently. 
As such, efforts were taken to release software that also incorporates ready-to-use machine learning models.
For example, Google released their \emph{MediaPipe} framework in 2019. \emph{MediaPipe} is an open-source framework developed by Google that provides a comprehensive solution for building machine learning-based applications for various multimedia tasks. It is designed to facilitate the development of applications that involve real-time processing of audio, video, and other sensor data \cite{lugaresi2019mediapipe}. \emph{MediaPipe} also includes a set of pretrained ML models, e.g., models for object detection, image segmentation or face detection.
Microsoft developed the \emph{Platform for Situated Intelligence} (\emph{psi}) \cite{bohus2021platform} - a versatile and open framework designed for development of integrative AI systems that leverage multi-modal capabilities while also incorporating various ML models for tasks such as voice activity or mouth shape detection.

Although those systems offer a plenitude of possibilities, they are, due to their industry-driven origin, designed to target a broad range of developers. 
As such, when building an application for a specific use case, still much effort has to be invested.
Besides that being a time-consuming factor, proficient programming knowledge is required to build the respective applications, which often can be a hindering factor - especially if the system is to be used in research projects where no computer science background is represented at all.

\section{System Architecture}
\label{sec:system}

\subsection{Overview}
The \emph{AffectToolbox} consists of a variety of independent components and modules that communicate to each other in a queue-based, multi-threaded runtime (Figure \ref{fig:system}).
Depending on how the components interact with the queue system, we categorize them into:\newline
\begin{itemize}
    \item \textbf{Input components:} Receive data from outside the framework (e.g., camera, microphones) and write it into at least one queue.
    \item \textbf{Processing components:} Read data from at least one queue, process it in various ways, and write new data in at least one queue (e.g., preprocessing algorithms, machine learning models, etc.)
    \item \textbf{Output components:} Read data from at least one queue, but as a data sink do not write new data to internal queues (e.g., GUI components, network adapters, etc.)\newline
\end{itemize}

As such, components can be easily added by defining an input queue that the new component should load its data from, and/or an output queue that the new component should write its data to, as well as the functionality that the component should serve.
Note that each module can process data in its own frequency - the respective frequency and other component-specific parameters are interfaced in a way that they can be adjusted via code or directly in the graphical user interface (GUI).

\begin{figure}[h]
\centering
\includegraphics[width=0.4\textwidth]{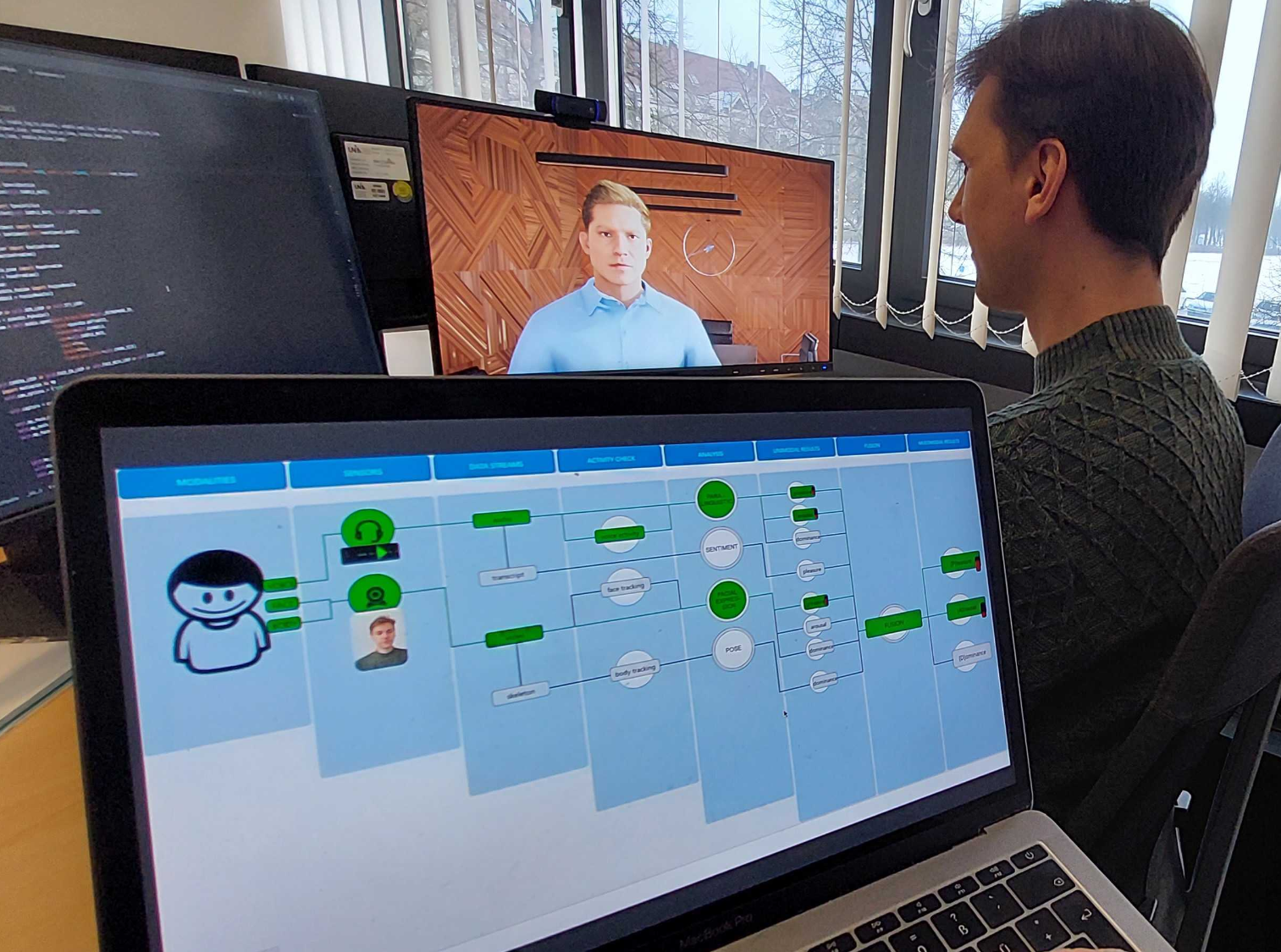}
\caption{User interacting with a virtual agent, whose emotional behaviour is taking real-time input from the \emph{AffectToolbox} into consideration.}
\label{fig:GUI}
\end{figure}

\subsection{Interface Design}
The GUI design of the \emph{AffectToolbox} is crucial for its accessibility across fields. 
We have carefully designed it in a way that basic and important functions are easily visible, avoiding confusion for new users by keeping unnecessary details out of the main design. 
However, note that all necessary configuration parameters can be adjusted in respective context menus in greater detail - as such, addressing the needs for slightly more sophisticated use cases as well.
In the GUI, all modalities and potential data flows are always on displays. 
Specific modalities can be activated with a simple click, triggering all necessary connections and dependencies automatically.
Visually highlighting the active data flow ensures that users see ongoing interactions in real-time.
The GUI can be seen in Figure \ref{fig:GUI}, where the \emph{AffectToolbox} is used to steer the emotional behaviour of a virtual agent.

\subsection{Emotion Model}
\label{sec:emo}

For the implementation of the \emph{AffectToolbox}, a common representation of the emotional user state has to be chosen. All included classification models need to recognize towards a respective affective golden standard. Also, the result is communicated to the user and / or subsequent applications in the chosen format.

%Defining a discrete representation of emotional states is a non-trivial task, as classifying emotional observations is not as objective as e.g. tagging a picture with its content.

%To narrow down the problem, we look at three terms which are commonly used in emotion research: Feelings, emotions and affect \cite{Damasio2003}. Feelings describe an internal mental state that is not depicted in bodily expressions whereas the term emotions is commonly used to describe the physiological manifestation of these feelings. Affect subsumes both phenomena under a general term, describing the cause as well as the portrayal of human emotions. Within the field of automatic affect recognition primary interest is given to the recognition of the mentioned portrayed affective states, the terms affect and emotion are mainly used as synonyms.

A common choice for automatic affect recognition are pre-selected categories or labels. Categorical emotion models subsume affective states under discrete categories like happiness, sadness, surprise or anger. This bears the advantage, that there is a wide and common understanding of these discrete categories of basic emotions \cite{Ekman1992}.

Usage of discrete labels is however restricting, as many blended feelings and emotions cannot adequately be described by the chosen categories. A more precise way to describe emotions is to describe the experienced stimuli with the help of continuous scales within dimensional models: Lang et al. \cite{Lang1997} suggest to characterize emotions along two continuous axes, pleasure and arousal. Mehrabian \cite{Mehrabian1995} proposes dominance as an additional measurements (\emph{PAD} model). The pleasure scale describes the pleasantness of a given emotion. A high value indicates an enjoyable emotion such as joy or happiness, lower values are associated with negative emotions like sadness and fear. The arousal scale measures the agitation level of an emotion. Dominance further refines the model by adding assessment of the feelings of being in control of a situation as well as autonomy.

\begin{figure}[h]
\centering
\includegraphics[width=0.35\textwidth]{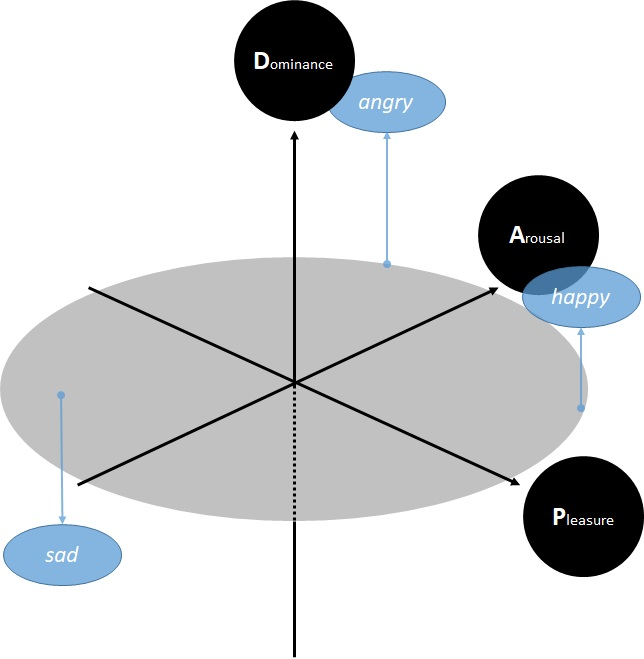}
\caption{The three dimensional \emph{PAD} model of affect. Continuous pleasure, arousal and dominance scores describe affective states, from which discrete emotion labels can be derived.}
\label{fig:emotion_model}
\end{figure}

These continuous representations are less intuitive but allow continuous blending between affective states. They describe multiple aspects of an emotion, the combination of stimuli's alignments on these scales defines single emotions (Figure \ref{fig:emotion_model}), which makes a conversion from dimensional values to emotional labels possible. These characteristics make the dimensional \emph{PAD} model our choice for handling the affective golden standard within the \emph{AffectToolbox}.

% Categorical as well as dimensional models are simplified and synthetic describtions of human affect and are not able to cover all of the included aspects. They are however useful and needed to model emotions as concepts to be presented to a machine. In addition to the categorical and dimensional models, appraisal models are in use to simulate the emotional behaviors of virtual agents in dialogue systems (Bee et al. \cite{Bee2010}). Among the most known examples are EMA (Gratch and Marsella \cite{Gratch2004}) and ALMA (Gebhard \cite{Gebhard2005}). There are only a few approaches that rely on appraisal models for affect recognition tasks (see Bosma and Andr\'e \cite{Bosma2004} for an early example). More recent work by Mortillaro et al. \cite{Mortillaro2012} suggests to infer emotions based on autonomic symptoms and motor behaviour as appraisal results. A first approach into this direction has been made by Soleymani \cite{Soleymani2016} with the detection of appraisal components, such as novelty, from facial expressions.

\subsection{Modalities}
\label{sec:modalities}

The \emph{AffectToolbox} applies signal processing and deep learning methods to interpret recorded signals by learning a mapping between observed signals and affective states. In uni-modal affect classification, information from one social channel,e.g. vocal properties, are used to make assumptions about the current emotional condition of a user. But as the cues that describe emotional conditions are indeed encoded within multiple modalities, the classification process should incorporate as much multi-modal information as possible \cite{Zeng2009}.

\subsubsection{Face Analysis}

Facial expressions are considered the most expressive transmitter of human emotions. Concentration and training would be needed to mask the depiction of affect in one's face reliably. But even then it takes a certain amount of time to control the muscle reactions that were triggered by the underlying emotional state and the correct emotion is expressed during this short period \cite{Ekman2009}. Analysis of facial expressions has greatly benefited from advances in image processing and resulted in reliable affect recognition models.

\paragraph{Camera Component}
Being able to connect to a variety of different camera hardware setups is essential for the accessibility of the toolbox.
As such, we rely on OpenCV's camera interface solutions.\footnote{https://opencv.org/}
By doing so, we ensure compatibility with the majority of existing hardware, as such supporting both standard and specialized use cases. According to a specified sample rate, our camera component takes an image frame and writes it to the respective input queue.

\paragraph{Face Preprocessing}
Before passing the camera data on to our machine learning models, we process them with two preprocessing components. First, we use the MediaPipe\footnote{https://github.com/google/mediapipe} face detection solution to extract the users' face. If a face is found, a bounding box is placed around it, which is used to crop the image such that only face-relevant information remains. 
If no face is found, the image is passed on as-is.
Subsequently, we normalize the images with min-max normalization ($min=-1.0, max=1.0$) and resize them to $224x224$ pixel.

\paragraph{Face Activity Detection}
We apply MediaPipe's face mesh detection solution to the preprocessed camera data.
The resulting face landmarks have two major benefits:
\begin{itemize}
    \item First, if no face mesh is detected at all, it is likely that no user is in the tracking area of the camera. Having that information, we can ignore the face modality for the respective timeframes.
    \item Second, by analyzing the landmark coordinates, we can obtain information on whether the user is actually facing the camera and/or screen. Depending on the application context, we can use that information to appropriately downweigh the face modality.
\end{itemize}

\paragraph{Deep Facial Analysis}
For the analysis of facial expressions we use a custom model that is based on the MobileNetV2 architecture \cite{sandler2018mobile}. Its modifications include the addition of two model heads to enable the recognition of continuous valence/arousal values and the detection of eight discrete emotion classes (\emph{Neutral, Happy, Sad, Surprise, Fear, Disgust, Anger} and \emph{Contempt}). The reasoning behind these choices was to improve the recognition accuracy while maintaining a lightweight architecture for faster processing \cite{schiller2020relevance}. During the training procedure, both tasks were learned simultaneously from the AffectNet dataset \cite{mollahosseini2019affect}. Afterwards, the secondary model head for discrete emotion detection was discarded as it only served to stabilize the primary task. The resulting model uses the preprocessed camera data as input and provides predictions for valence/arousal values as output.

\begin{figure}[h]
\centering
\includegraphics[width=0.45\textwidth]{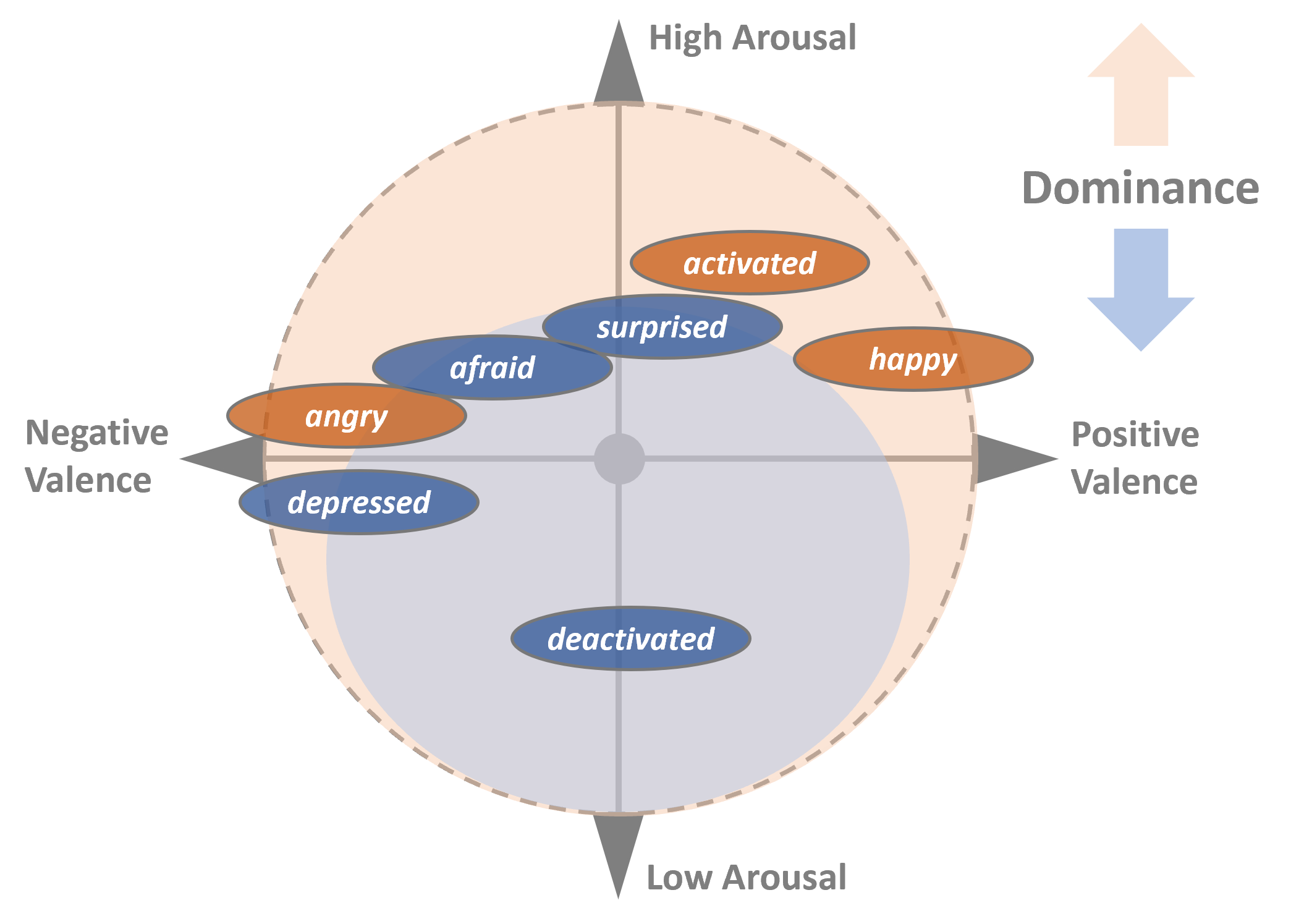}
\caption{Based on empirical studies, a dominance level can be derived from facial expressions described in the valence/arousal space or inferred emotional labels.}
\label{fig:dominance_face}
\end{figure}

While there exist public corpora with reliable valence and arousal annotations, the dominance dimension is less covered by respective resources. To also derive a dominance score from the facial modality, we chose a literature-based approach: Studies such as \cite{russell1977evidence} describe observed relations between the three scales of the PAD model as well as emotional labels. Based on these insights, one can infer a perceived dominance level from recognized facial expressions described in the valence/arousal space (or inferred emotional label).

%The structure of that model was based on the MobileNetV2 architecture \cite{Sandler2018}, which was adapted for multi-task learning of the following two tasks: recognition of continuous valence/arousal values and detection of eight discrete emotion classes (\emph{Neutral, Happy, Sad, Surprise, Fear, Disgust, Anger} and \emph{Contempt}). We selected this architecture because it achieves similar results to Inception- or ResNet-based models but can be trained more quickly \cite{schiller2020relevance}.
%The network was trained for both tasks simultaneously on the AffectNet dataset \cite{affectnet} using an \emph{Adam} optimizer with a learning rate of 0.001 for 100 epochs. After training, the secondary model head for discrete emotion detection was removed and only the valence/arousal head was used.

%\bgroup
%\newcolumntype{Y}{>{\centering\arraybackslash}X}
%\newcolumntype{Z}{>{\raggedleft\arraybackslash}X}
%\begin{table}[htb]
%	\centering
%	\begin{tabularx}{\columnwidth}{@{}XYYYY@{}}
%		\toprule[\lightrulewidth]
%		& \multicolumn{2}{c}{AffectNet Baseline} & \multicolumn{2}{c}{Evaluation Model}\\
%		\cmidrule(l){2-3}
%		\cmidrule(l){4-5}
%		& Valence & Arousal & Valence &  Arousal \\
%		\midrule[0.1em]
%		RMSE & 0.37 & 0.41 & 0.40 & 0.37\\
%		CORR & 0.66 & 0.54 & 0.60 & 0.52\\
%		SAGR & 0.74 & 0.65 & 0.73 & 0.75\\
%		CCC  & 0.60 & 0.34 & 0.57 & 0.44\\
%		\bottomrule[\lightrulewidth]
%		\addlinespace[\belowrulesep]
%	\end{tabularx}
%	\caption{AffectNet performance comparison}
%	\label{tbl:affectnet-performance}
%\end{table}
%\egroup

\subsubsection{Pose Analysis}

Although early work like Caridakis et al. \cite{Caridakis2006A} describes the possibilities of expressivity features (mainly calculated on arm and head movements extracted from video sequences), the main assumption was that body movement only shows the intensity of emotions. However, in recent years studies have shown that dynamic body movement and gestures as well as static postures convey affective states of a monitored person. Affective concepts such as a person's current expressivity \cite{Caridakis2010} or engagement in a conversation \cite{Baur2016} can be reliably described by suitable movement features, body orientation, and posture. Spatial expansive poses and postures including e.g. a straightened spine are used to convey dominance and self-confidence and (though a problem of replicating respective studies persists) there are clear hints towards effects on self perception and emotional experience \cite{koerner2020}.

% The actions and positions of of body, head and limbs - also referred to as \emph{kinesics} - have for a rather long time not been the main focus of coordinated emotion research

%The trend to more accurately analyse gestures and postures as a mean to affect recognition was most likely positively affected by technical advancements like the Microsoft Kinect\texttrademark, that make whole body tracking a more accessible approach. By 2012 a comprehensive survey by D'Mello and Kory \cite{DMello2012} investigated 30 current multi-modal affect recognition systems and stated that almost a third of these systems were by then using information from some form of body movement, postures and gestures as a source for emotion assessments.

\paragraph{Skeleton Detection}

The actions and positions of body, head, and limbs - also referred to as \emph{kinesics} - can be derived from detected pose landmarks in images. As a reliable detection model, we chose to integrate the lightweight convolutional neural network BlazePose \cite{bazarevsky2020blazepose}, with which we can define a skeleton overlay for each video frame to be further processed into pose features.

\paragraph{Pose Features}

A logical consequence of this rising interest in kinesics is the effort to develop a reliable coding system for body movement in emotion expression, just like the by-now-available standards for facial expressions. For this reason, Dael et al. \cite{Dael2012B} describe the Body Action and Posture System: The system distinguishes body posture units and body action units \cite{Harrigan2008} with the first representing the general alignments of trunk, head and limbs to a resting configuration (e.g. arms crossed) and the latter one describing a local and short termed movement of head or arms (e.g. pointing gesture). We went for a comparable rule-based approach by deducing a perceived dominance score based on pose features calculated from the detected pose landmarks (skeleton). Features by now include head and body tilts as well as overall activation of a user. Further features are continuously developed.

\subsubsection{Voice Analysis}
\label{sec:va}
\begin{figure}[h]
\centering
\includegraphics[width=0.4\textwidth]{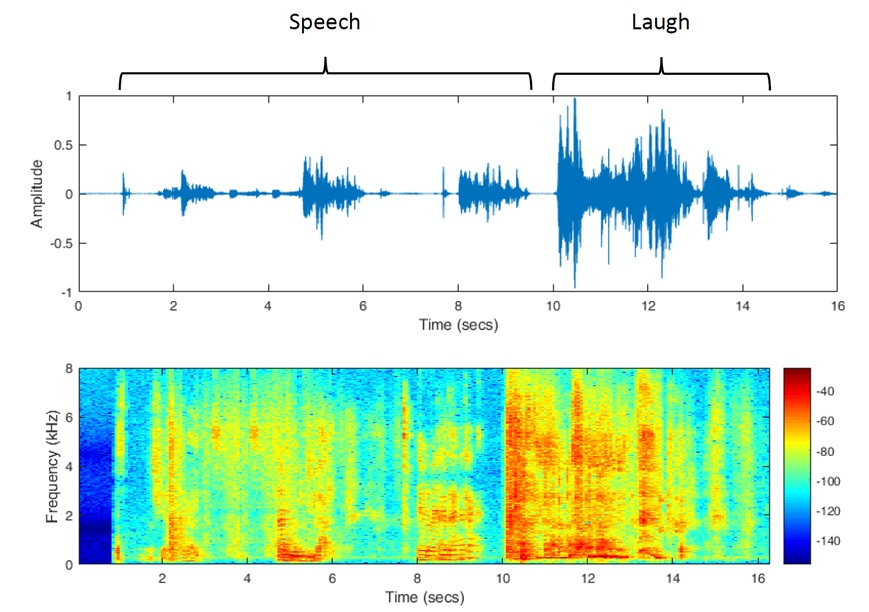}
\caption{Audio signal of a person's normal speech leading into an affective burst of laughter. The characteristics of the signal parts can be well differentiated. Further processing of the signal carves out detailed differences and enables the categorization of more subtle affective states.}
\label{fig:audio}
\end{figure}

Human language encodes emotional information in a semantic as well as in a paralinguistic way \cite{Wagner2015}. Paralinguistics refers to phenomena that accompany speech and do not consist of linguistic units such as sounds, words, sentences, etc., but give it an additional communicative aspect. This includes acoustic characteristics such as pitch, volume or speaking speed.Figure \ref{fig:audio} shows the audio signal of a person's normal speech leading into an affective burst of laughter. The characteristics of the signal parts can be well differentiated. 
Semantics on the other hand describes the content and the grammatical format of utterances, including the arrangement and choices of words, phrases, and clauses.
Although modern language analysis models show promising results in analyzing both aspects of language within a unified architecture \cite{wagner2023dawn}, it is still advantageous to analyze paralinguistics and semantics separately for optimal performance and flexibility.

\paragraph{Microphone Component}
To track the user's speech, we use PyAudio\footnote{https://github.com/CristiFati/pyaudio}. We chose PyAudio because it enables us to directly connect to a multitude of different hardware setups and allows for a large variety of different technical configuration options. The tracked audio data is written into the respective input queue in chunks of predefined size. As such, although we can keep the communication frequency with the microphone comparably low, we still have the whole audio signal in the queue system without any quality losses. By tweeking the chunk size and grabbing frequency, the application-dependent trade-off between low latency and resource efficiency can be controlled if necessary.

\paragraph{Voice Activity Detection}
For affective voice analysis, it is crucial to know when the tracked audio signals contain actual speech and not only background noises. As such, we make use of the WebRTC voice activity detector\footnote{https://github.com/wiseman/py-webrtcvad}. We feed the audio by using a sliding window, resulting in continuous probability estimations for the presence of voice. We can use those probabilities to downweigh or even completely ignore all voice-related modalities.

\paragraph{Paralinguistic Analysis}
Paralinguistic analysis refers to the analysis of voice properties unrelated to speech's linguistic content. As such, characteristics like intonation, pitch or loudness play an important role here. However, in recent years, performing paralinguistic analysis in an end-to-end manner, i.e., without handcrafting specific features, is becoming more and more standard\cite{wagner2018deep}. As such, for the \emph{AffectToolbox}, we use a current state-of-the-art network architecture for assessing pleasure, arousal, and dominance from voice\cite{wagner2023dawn}. The model is based on the popular wav2vec2 architecture\cite{baevski2020wav2vec} and has proven to obtain high performance on standard emotion recognition benchmark tasks\cite{wagner2023dawn}.

\subsubsection{Sentiment Analysis} 
%Deutsch/Englisch, checken ob englisch verfügbar
When analyzing emotions from spoken language, the paralinguistic analysis of speech is complemented by the semantic analysis of the spoken content. 
While the employed voice analysis model (\ref{sec:va}) is already incorporating linguistic information to some extent \cite{Triantafyllopoulos2022}, we argue that the integration of a dedicated sentiment analysis module is still adding value to the overall pipeline.

\paragraph{Speech-To-Text}
In the preceding step, we must first convert the spoken language into a textual representation that can be processed by a sentiment analysis model.  
To this end we rely on the \textit{whisper} model suite by Radford et el. \cite{Radford2023robust}
The model suite consists of an off-the-shelf encoder-decoder Transformer architecture as proposed by \cite{vaswani2017attention} at different scales. 
The novelty of the proposed work lies in the introduction of a weakly supervised multitask training procedure that relies heavily on large-scale data crawled from the web. 
The authors found that their proposed training procedure enables the model to perform many speech processing pipeline tasks like \textit{multilingual speech recognition}, \textit{speech translation}, \textit{spoken language identification}, and \textit{voice activity detection} at the same time while also enabling the resulting models to be competitive with other supervised models but in a zero-shot transfer setting, i.e. on data for which the model has not seen any examples.
All in all, these features make the model an ideal fit for the \emph{AffectToolbox}.

\paragraph{Sentiment}
To analyze the sentiment of a spoken utterance we rely on the XLM roBERTa model from Barbieri et al. \cite{barbieri2022xlm}
The roBERTa model is a pretrained variant of the BERT \cite{Devlin2018bert} architecture introduced by Liu et al. \cite{liu2019roberta} that relies upon an optimized training procedure to improve results.
The XLM roBERTa multilingual model has been further trained on a vast amount of Twitter data and provides support for sentiment analysis in eight different languages.
The authors chose the model architecture and datasets specifically to make it suitable for agent-human-interaction scenarios, which is also an ideal scenario for the \emph{AffectToolbox}

%To analyze the sentiment of a spoken utterance we rely on the GermanSentiment model from Guhr et al. \cite{Guhr2020}. 
%This model is a variant of the BERT architecture introduced by Devlin et al. \cite{Devlin2018bert} that has been trained on a combination of various existing datasets.
%The authors chose the model and datasets specifically to make it suitable for agent-human-interaction scenarios.

% https://huggingface.co/cardiffnlp/twitter-xlm-roberta-base-sentiment

\subsection{Fusion}
\label{sec:fusion}

Including multiple modalities in the affect recognition process is generally meant to enhance performance. In addition to the increase in information, the system as a whole becomes more robust: Modalities that temporarily - due to tracking failures or general lack of activity - do not contribute meaningful information can be substituted by other channels. The most impactful factor for the quality of a multi-modal affect recognition system is the ability to extract informative features and results from its single modalities (Section \ref{sec:modalities}). The fusion strategy to integrate this information into a coherent decision is however of equal importance for the design of the \emph{AffectToolbox}.\newline

\subsubsection{Requirements}

A key distinction between considered fusion approaches relates to the modeling of temporally shifted occurrences of emotional cues throughout multiple affective channels. While simple fusion strategies mainly apply a synchronous strategy, which considers all information within a fixed time segment, more elaborate systems regard these asynchronous characteristics of emotional manifestations and try to model them within the fusion process. Multi-modal affective events, such as facial expressions or vocal bursts that manifest certain emotional states, are expected to occur at shifted points in time and we consequently need to asynchronously treat modalities. We can therefore formulate key requirements for the \emph{AffectToolbox} fusion algorithm.\newline

\textbf{Temporal Flow}: If we recognize an affective cue in a modality, it enters the fusion process and influences the continuous result. The initial influence will diminish over time until the cue is ´no longer relevant and gets discarded. Current cues therefore have a stronger impact on the fusion process than the ones that lie further down the time axis.

\textbf{Reinforcement and Attenuation}: If complementary cues are detected in overlapping time-segments, they reinforce each other by amplifying their impact on the continuous fusion output. On the other hand, contradictory cues neutralize each other and therefore have a lesser negative and attenuating effect on the fusion result. This way, additional information from multiple modalities is more likely to enhance the overall classification performance.

\textbf{Real-Time Fusion Result}: The \emph{AffectToolbox} targets applications, that are reactive to current emotional states of a user. We consequently aim for near real-time recognition tasks on online input, and need the processing speed of all all algorithms to be suitable for this task. The result of the fusion scheme is calculated by temporal influences (expressed through diminishing weights) of registered cues and a current fusion result has to be available at any given point in time, guaranteeing access to the latest affective estimation for all subsequent components of an application.\newline

\begin{figure}[h]
\centering
\includegraphics[width=0.35\textwidth]{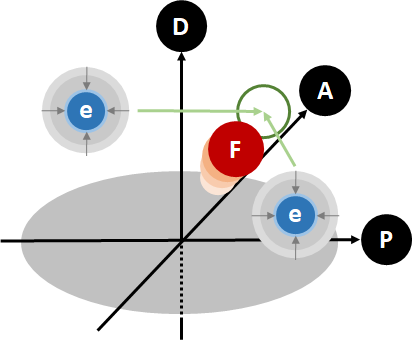}
\caption{The fusion result is calculated from active events in fixed update steps: At each time frame, the influence of active event vectors is reduced based on the defined decay speed, expired lifetime, and the initial norm of the vector and weight of the event. Within the three-dimensional PAD space, the fusion result is drawn towards the center of mass specified by weighted event vectors.}
\label{fig:fusion}
\end{figure}

\subsubsection{Implementation}

We based the implementation of our fusion mechanism on preceding work done by Gilroy et al. \cite{Gilroy2009}, which represents emotions as a vector within a dimensional emotion model. We generalized this approach by designing an event-driven fusion scheme that operates in a user-defined vector space \cite{Lingenfelser2014}, \cite{Lingenfelser2016}.

Affective cues, recognized in observed modalities, are registered as events for an event-driven algorithm, which continuously calculates its multi-modal fusion result based on the currently active affective events (Figure \ref{fig:fusion}). As the \emph{AffectToolbox} is operating with the pleasure-arousal-dominance emotion model, respective events are represented as three-dimensional PAD vectors and provided with several parameters: A score, which defines the position of the vector within the dimensional PAD model, is assigned for each axis in the event space. Each vector is given a weight parameter, which serves as a quantifier for its impact on the calculation of the fusion result. This way reliability of a modality or cue type can regulated. Finally, a decay speed parameter describes the average lifespan of cues extracted from the respective signal. It determines the time it takes for the event's influence to fully diminish and get discarded. Events with strong indications can be given longer decay times to prolong their influence on the result.\newline

%[Score] This value can be directly given by the continuous output of a classification model or can be dynamically calculated from the normed probabilities of a recognized cue, resulting in values that typically range between 1 and -1.

Given these parameters, the fusion result can be calculated from active events at each update step: At each time frame, active event vectors $e=1\ldots E$ are decayed by multiplying each vector element with a decay factor that is calculated based on the defined decay speed, expired lifetime and the initial norm of the vector:
\begin{equation}
decay_{e} = norm_{e} - (lifetime_{e} * speed_{e})
\end{equation}
If the resulting norm of the decayed vector stays above zero, it remains active - otherwise the vector is discarded. Afterwards, the fusion point within the vector space is calculated from all active event vectors: For each dimension $d=1\ldots D$ of the vector space respective scores of active event vectors $e=1\ldots E$ (modified by their weight factor) are summed up.
\begin{equation}
fusion\:point(d_{1\ldots D}) = \sum_{e=1}^{E}{(event\:vector_{d,e} * weight_{e})}
\end{equation}
The result is normalized by the sum of the weights of all contributing event vectors.
\begin{equation}
fusion\:point(d_{1\ldots D}) = mass(d_{1\ldots D}) / \sum_{e=1}^{E}{weight_{e}}
\end{equation}
The final result itself is a vector that approaches the calculated fusion point within the PAD space with a predefined speed parameter (Figure \ref{fig:fusion}). If no events remain active in the vector space, the fusion vector approaches a neutral state. The fusion vector serves as an additional means of smoothing: If we can assume that the thought affective state is unlikely to undergo quick changes, the vector can be defined to move slowly, making the algorithm more robust to occasional misclassifications.

%Also note that the calculation of the fusion point leads to axial dependencies in a multidimensional event space. This may be of interest in a scenario where the positive recognition of a class is meant to have a decreasing effect on the likelihood of another one. If axial independence is needed, the use of several one-dimensional models with subsequent combination is advised.

\section{Discussion}
We have by now seen how the \emph{AffectToolbox} can analyze multi-modal behaviour and subsequently deduce the affective states of users. This capability offers a wide range of possible applications but also features some technical requirements and conceptual limitations that have to be taken into account.

\subsection{Potential Use-Cases}
Monitoring of affective user states is of great interest in theoretical research scenarios as well as practical software design. Psychological studies often include automated analysis and recording of proband behaviour. Several instances of the \emph{AffectToolbox} can even be used to track emotional states of several study participants (e.g. dyadic or multi-person conversations, therapist and patient, etc.).

In most applications for which the \emph{AffectToolbox} was specifically designed, the emotional analysis is used in a human-computer interaction (HCI) scenario, i.e. the interaction between human users and a virtual agent. User emotion can hereby serve as contextual information for the dialogue system to also include the affective intent of the user in steering the conversational flow. Also, the non-verbal behaviour of a virtual character can be designed in a very natural and reactive manner if insights into the current emotional condition of the human counterpart are available.

A further use-case in the field of HCI is the adaptation of systems to user states. Techniques such as reinforcement learning can use e.g. PAD values as additional input to the knowledge space or in the calculation of the reward function. This way implicit feedback mechanisms greatly enhance the user experience.

In all described fields of application, study and system designers can benefit greatly from a comprehensive analysis tool, that is easy to set up and integrate into the broader infrastructure without the need to allocate resources to the implementation of the affect recognition pipeline. To this end, the \emph{AffectToolbox} includes networking capabilities to broadcast its results in several formats (JSON, XML, etc.) and protocols (UDP, Kafka, etc.).

%- Studies (Psychologischer Bumms, Zustände von Probanden tracken) Recording
%- Dialogsteuerung - Verhaltenssteuerung (Agent/Roboter, z.B, Dialog oder Gestik) 
%- Systemadaption - Trainingssysteme (Schwierigkeitsanpassung, Informationsgehalt)

\subsection{Technical Requirements}
To hold on to the ease-of-use paradigm, we try to keep the technical requirements of the \emph{AffectToolbox} within reasonable limits: Audiovisual sensory devices (e.g. webcams) provide easy means to generate all so far considered data streams (i.e. audio, transcript, video, and skeleton data) without reliance on expensive or complicated sensor hardware.

The same approach is followed for our reliance on computing power. The \emph{AffectToolbox} features an extensive list of tuning parameters (e.g. analysis frame rates, variable model sizes, etc.) to customize the system to the capabilities of available hardware. Of course, the best performance with a high frequency and accuracy of classification results is achieved with an up-to-date hardware setup, we can nevertheless achieve acceptable results with less costly and more available configurations. This is partly due to the fusion process, which is inherently able to interpolate within a sparse result space (Section \ref{sec:fusion}).

%- low hardware requirements (sowohl sensor als auch hardware)
%- tuning parameters

\subsection{Limitations}
Generally we consider the \emph{AffectToolbox} to be able to recognize so-called \emph{displayed} emotions. This means, that - at least with the audiovisual information used - there is no reliable way to differentiate between a genuine feeling or socially altered affective state. Whether the displayed and respectively measured emotions are felt or expressed for other reasons, like using them as conversation techniques, resulting PAD values will be the same from the system's point of view.

Ways to investigate deeper into the actually felt emotion mostly include context information and interpretation or additional physiological sensory equipment. Physiological responses (e.g. heart rate, skin conductivity, and respiration as well as eye gaze and pupil dilation) are more unconscious reactions to affect that are hard to control by untrained subjects, as they are regulated by the autonomous nervous system. Inclusion of this kind of analysis would on the one hand contradict the accessability paradigm of the \emph{AffectToolbox}, on the other hand might be worth the effort to optionally offer further valuable user states such as stress, cognitive workload or user engagement (Section \ref{sec:conclusion}).

%- physiologischer shit, selected modalities atm
%- nur gezeigtes verhalten blabla

\section{Conclusion \& Future Work}
\label{sec:conclusion}
The \emph{AffectToolbox} is able to offer multi-modal recognition of displayed user emotions in real-time. Results are given in fine-grained continuous pleasure-arousal-dominance levels. In contrast to related solutions, it is easy to apply, as no programming knowledge is required: A concise graphical user interface provides accessible configuration and customization possibilities. High-quality and ready-to-use recognition models are integrated into the toolbox for each modality, a sophisticated fusion algorithm has been implemented to handle the multi-modal affective cues and generate a coherent result. Integration into application setups is easy to accomplish as fusion results are continuously broadcasted and accessible at any point in time.

Several research projects are currently using the \emph{AffectToolbox} for the development of affect-sensitive studies and prototypes. Given the multi-disciplinary nature of these groups, we see numerous non-technical users applying the toolbox without major complications. Acceptance of the system underlines and reinforces the goal to provide an easy-to-handle solution to integrate affective computing into state-of-the-art HCI systems.

The toolbox is in active development, which first means keeping the system up-to-date with novel signal processing methods and machine learning models. Second, we are investigating potential the demand for additional affective states beyond PAD values. Offering insights into more medical and well-being-related user states such as stress levels together with emotional states seems to raise interest in potential user groups. Automatic measurement of engagement in interactions as well as estimations of cognitive workload are expected to be of great benefit for the adaptation of HCI systems to current user needs, e.g. in training or educational scenarios.

% Furthermore alles geil, respectively weil is already in use in several research projects. qed ... hier Mithos etc. rein. Conclusion is dann quasi - ja wird benutzt
% future work: weitere affektive zustände, wie z.B, stress(fordigitstress paper), engagement(tobis diss), cognitive workload

\section*{Acknowledgments}
This research is funded by the Federal Ministry of Education and Research, Project MITHOS (grand agreement 16SV8687) and Project UBIDENZ (grand agreement 13GW0568F).

\newpage

\printbibliography

\end{document}